\documentclass[aps,prb,twocolumn,superscriptaddress]{revtex4-1}

\usepackage{graphicx}% Include figure files
\usepackage{dcolumn}% Align table columns on decimal point
\usepackage{bm}% bold math
\usepackage[version=3]{mhchem}

\begin{document}

\title{Defect chemistry in layered transition-metal oxides from screened hybrid density functional calculations}
\author{Khang Hoang}
\email[E-mail: ]{khang.hoang@ndsu.edu}
%\homepage[]{Your web page}
%\thanks{}
%\altaffiliation{}
\affiliation{Center for Computationally Assisted Science and Technology, North Dakota State University, Fargo, ND 58108, USA.}
\author{Michelle D.~Johannes}
%\email[]{}
%\homepage[]{Your web page}
%\thanks{}
%\altaffiliation{}
\affiliation{Center for Computational Materials Science, Naval Research Laboratory, Washington, DC 20375, USA.}

%\date{\today}

\begin{abstract}

We report a comprehensive first-principles study of the thermodynamics and transport of intrinsic point defects in layered oxide cathode materials \ce{LiMO2} (M=Co, Ni), using density-functional theory and the Heyd-Scuseria-Ernzerhof screened hybrid functional. We find that \ce{LiCoO2} has a complex defect chemistry; different electronic and ionic defects can exist under different synthesis conditions, and \ce{LiCoO2} samples free of cobalt antisite defects can be made under Li-excess (Co-deficient) environments. A defect model for lithium over-stoichiometric \ce{LiCoO2} is also proposed, which involves negatively charged lithium antisites and positively charged small (hole) polarons. In \ce{LiNiO2}, a certain amount of Ni$^{3+}$ ions undergo charge disproportionation and the concentration of nickel ions in the lithium layers is high. Tuning the synthesis conditions may reduce the nickel antisites but would not remove the charge disproportionation. In addition, we find that \ce{LiMO2} cannot be doped $n$- or $p$-type; the electronic conduction occurs via hopping of small polarons and the ionic conduction occurs via migration of lithium vacancies, either through a monovacancy or divacancy mechanism, depending on the vacancy concentration.

\end{abstract}

% insert suggested PACS numbers in braces on next line
%\pacs{}
% insert suggested keywords - APS authors don't need to do this
%\keywords{}

%\maketitle must follow title, authors, abstract, \pacs, and \keywords
\maketitle

% body of paper here - Use proper section commands
% References should be done using the \cite, \ref, and \label commands

\section{Introduction}

Layered transition-metal oxides \ce{LiMO2} (M=Co, Ni) have been studied intensively for lithium-ion battery intercalation cathodes.\cite{Whittingham_CR,Kalyani2005} It has been reported that \ce{LiCoO2} synthesized by conventional high temperature ($>$800$^\circ$C) procedures possesses the $O3$-type layered structure with excellent ordering of the Li$^{+}$ and Co$^{3+}$ ions and good structural stability. The material synthesized at low temperatures ($\sim$400$^\circ$C), however, yields a significant disordering of the Li$^{+}$ and Co$^{3+}$ ions and exhibits poor electrochemical performance.\cite{Gummow1992327} It has also been shown that experimental studies of the magnetic properties of \ce{LiCoO2} always reveal localized magnetic moments, and the commercially available, high-temperature synthesized \ce{LiCoO2} is often made deliberately with Li-excess.\cite{Chernova2011} The electronic state of Co and the nature of charge-compensating defects in \ce{LiCoO2} are still not fully understood, although several defect models for the lithium over-stoichiometric ({\it i.e.},~Li-excess) \ce{LiCoO2} have been proposed.\cite{Carewska1997,Levasseur2000,Levasseur2003,Hertz2008,Carlier2013} Regarding \ce{LiNiO2}, it is known that the compound with all Ni in a $3+$ valence state is extremely hard to synthesize because of the difficulty of stabilizing Ni$^{3+}$ at high temperatures.\cite{Kalyani2005,Dutta1992123,Hirano1995,Kanno1994216} In fact, it is now believed that stoichiometric \ce{LiNiO2} does not actually exist, and the compound has always been found to have a significant concentration of Ni ions at the Li sites.\cite{Barra1999,Chappel2002,Kalyani2005,Chernova2011} The off-stoichiometry and cation mixing have been found to be detrimental to \ce{LiNiO2}'s electrochemical performance.\cite{Kalyani2005,Whittingham_CR,Chernova2011}

In order to understand the experimental observations and resolve the conflicting defect models and interpretations, apparently one needs to have a detailed understanding of the materials' defect chemistry, which can be achieved through first-principles computational studies. In fact, calculations based on density-functional theory (DFT) have been proven to be an important tool in investigations of point defects in battery cathode materials. In such calculations, certain aspects can be isolated and studied more easily than in experiments. A proper and comprehensive treatment of defects based on DFT not only can provide a quantitative understanding of the thermodynamics and transport of the defects but also shed light on the electronic and ionic conduction mechanisms and help develop strategies for improving the materials' performance.\cite{hoang2011,Hoang2012274,Johannes2012}

There have been numerous computational studies of layered \ce{LiMO2}.\cite{anisimov1991,VanderVen2000,Kang2006,Wang2007,Laubach2009,Ensling2010,Chevrier2010,Chen2011} All these studies have, however, focused mainly on the bulk properties and Li diffusion, with much less attention paid to the defect chemistry. A systematic DFT study of point defects in \ce{LiMO2} has only recently been carried out by Koyama {\it et al.},~\cite{Koyama2012,Koyama2013} providing useful information on defect formation in the studied materials. This work, however, has three major limitations. Firstly, the DFT calculations were carried out using the GGA$+U$ method,\cite{dudarev1998,anisimov1991,anisimov1993,liechtenstein1995} an extension of the generalized-gradient approximation (GGA) for the exchange-correlation functional,~\cite{GGA} in which the on-site Coulomb interaction parameter $U$ for the transition metal was assumed to be the same in different chemical environments. Secondly, the authors did not address the spurious long-range Coulomb interactions between charged defects in calculations using the supercell approach.\cite{walle:3851,Freysoldt,Freysoldt11} These interactions often significantly alter the calculated total energies, leading to inaccurate defect formation energies. Thirdly, and most importantly, the authors did not fully investigate the dependence of defect formation energies on the atomic chemical potentials, which can be used to represent the synthesis conditions, resulting in an incomplete and inaccurate picture of the defect chemistry.

In this article, we present a comprehensive DFT study of the structure, energetics, and migration of intrinsic point defects in layered \ce{LiMO2} using a hybrid Hartree-Fock/DFT method, specifically the Heyd-Scuseria-Ernzerhof (HSE06) screened hybrid functional.\cite{heyd:8207,paier:154709,heyd2006} Compared to the GGA$+U$ method, the hybrid functional improves the transferability of calculations across compounds by treating all orbitals on the same footing, thus improving the accuracy of defect formation energies. We find that \ce{LiCoO2} has a surprisingly complex defect chemistry. Different electronic and ionic defects such as small polarons, antisite defects, and lithium vacancies can exist with high concentrations in the material when synthesized under different conditions. In \ce{LiNiO2}, a certain amount of Ni$^{3+}$ ions undergo charge disproportionation, and nickel antisites have a low formation energy and hence high concentration. We will discuss how our results for \ce{LiMO2} can explain the experimental observations, help understand the mechanisms for electronic and ionic conduction, assist in defect characterization and defect-controlled synthesis and ultimately aid in the rational design of cathode materials with improved electrochemical performance. Comparison with previous theoretical works will be made where appropriate.

\section{Methodology}
\subsection{Computational details}

The presented calculations were based on DFT using the HSE06 hybrid functional,~\cite{heyd:8207,paier:154709,heyd2006} and the projector augmented wave method,~\cite{PAW1,PAW2} as implemented in the VASP code.~\cite{VASP1,VASP2,VASP3} The GGA+$U$ method~\cite{dudarev1998,anisimov1991,anisimov1993,liechtenstein1995} was used only for comparison in some specific bulk calculations, with $U$ values set to 4.91 eV (for Co) and 6.70 (Ni), taken from Zhou {\it et al.}~\cite{Zhou:2004p104} Calculations for bulk \ce{LiMO2} in the $O3$-type layered structure were performed using a 7$\times$7$\times$7 Monkhorst-Pack $\mathbf{k}$-point mesh.\cite{monkhorst-pack} The structural optimization allowed for Jahn-Teller distortion in \ce{LiNiO2}. Intrinsic point defects were treated within the supercell approach, in which a defect is included in a finite volume of the host material and this structure is periodically repeated. For defect calculations, we used hexagonal (3$\times$3$\times$1) supercells, which correspond to 108 atoms/cell; integrations over the Brillouin zone were carried out using the $\Gamma$ point. The plane-wave basis-set cutoff was set to 500 eV. Convergence with respect to self-consistent iterations was assumed when the total energy difference between cycles was less than 10$^{-4}$ eV and the residual forces were less than 0.01 eV/{\AA}. In the defect calculations, the lattice parameters were fixed to the calculated bulk values, but all the internal coordinates were fully relaxed. The migration of selected defects in \ce{LiMO2} was studied using the climbing-image nudged elastic-band (NEB) method.\cite{ci-neb} All calculations in \ce{LiMO2} were performed with spin polarization and the ferromagnetic spin configuration.

\subsection{Defect formation energies}

The formation energy $E^f$ of a defect is a crucial factor in determining its concentration. In thermal equilibrium, the concentration at temperature $T$ can be obtained via the relation~\cite{walle:3851} 
\begin{equation}\label{eq:concen} 
c=N_{\mathrm{sites}}N_{\mathrm{config}}\mathrm{exp}\left(\frac{-E^{f}}{k_{B}T}\right), 
\end{equation} 
where $N_{\mathrm{sites}}$ is the number of high-symmetry sites in the lattice per unit volume on which the defect can be incorporated, and $N_{\mathrm{config}}$ is the number of equivalent configurations (per site). It follows from this equation that defects with low formation energies will easily form and occur in high concentrations.

The formation energy of a defect X in charge state $q$ is defined as~\cite{walle:3851,hoang2011}
\begin{eqnarray}\label{eq:eform}
\nonumber
E^f({\mathrm{X}}^q)=E_{\mathrm{tot}}({\mathrm{X}}^q)&-&E_{\mathrm{tot}}({\mathrm{bulk}})-\sum_{i}{n_i\mu_i} \\
&+&q(E_{\mathrm{v}}+\mu_{e})+ \Delta^q ,
\end{eqnarray}
where $E_{\mathrm{tot}}(\mathrm{X}^{q})$ and $E_{\mathrm{tot}}(\mathrm{bulk})$ are, respectively, the total energies of a supercell containing the defect X in charge state $q$ and of a supercell of the perfect bulk material; $\mu_{i}$ is the atomic chemical potential of species $i$ (and is referenced to its standard state), and $n_{i}$ denotes the number of atoms of species $i$ that have been added ($n_{i}$$>$0) or removed ($n_{i}$$<$0) to form the defect. $\mu_{e}$ is the electronic chemical potential, referenced to the valence-band maximum in the bulk ($E_{\mathrm{v}}$). $\Delta^q$ is the correction term to align the electrostatic potentials of the bulk and defect supercells and to account for finite-cell-size effects on the total energies of charged defects.\cite{walle:3851} To correct for the finite-size effects, we adopted the approach of Freysoldt {\it et al.}~in which the correction term to the formation energy is determined without empirical parameters.\cite{Freysoldt,Freysoldt11} This approach has proven to be effective for studies of defects in solids.\cite{Hoang201353,Komsa2012}

The atomic chemical potentials $\mu_{i}$ are variables and subject to thermodynamic constraints, which can be used to represent the synthesis conditions.\cite{walle:3851,hoang2011} The stability of \ce{LiMO2} requires
\begin{equation}\label{eq;stability} 
\mu_{\rm Li}+\mu_{\rm M}+2\mu_{\rm O}=\Delta H^{f}({\rm LiMO}_{2}), 
\end{equation} 
where $\Delta H^{f}$ is the formation enthalpy. This condition places a lower bound on the value of $\mu_{i}$. Additionally, one needs to avoid precipitating bulk Li and M, or forming O$_{2}$ gas. This sets an upper bound on the chemical potentials: $\mu_{i}$$\leq$0.\cite{walle:3851} In our work, the zero reference state of $\mu_{\mathrm{O}}$ is chosen to be half of the total energy of an isolated O$_{2}$ molecule at 0 K.

There are further thermodynamic constraints imposed by competing Li$-$M$-$O phases which usually place stronger bounds on $\mu_{i}$. For example, in order to avoid the formation of Li$_{2}$O, a competing phase of \ce{LiCoO2},
\begin{equation}\label{eq;li2o} 
2\mu_{\rm Li}+\mu_{\rm O}\leq \Delta H^{f}({\rm Li}_{2}{\rm O}). 
\end{equation}
By taking into account the constraints imposed by all possible competing phases, one can define the range of Li, M, and O chemical potential values in which \ce{LiMO2} is stable.

The electronic chemical potential $\mu_{e}$, hereafter also referred to as the Fermi level, is not a free parameter either. In principle, eqns~(\ref{eq:concen}) and (\ref{eq:eform}) can be written for every intrinsic defect and impurity in the material. The complete problem, including free-carrier concentrations in valence and conduction bands, if present, can then be solved self-consistently by imposing the charge neutrality condition:\cite{walle:3851}
\begin{equation}\label{eq:neutrality}
\sum_{i}c_{i}q_{i}-n_{e}+n_{h}=0,
\end{equation}
where $c_{i}$ and $q_{i}$ are the concentration and charge of defect or impurity X$_{i}$; $n_{e}$ and $n_{h}$ are free electron and hole 
concentrations; the summation is over all defects and impurities.

\section{Results}
\subsection{Bulk properties}

\begin{figure}
\vspace{0.2cm}
\centering
\includegraphics[height=5.8cm]{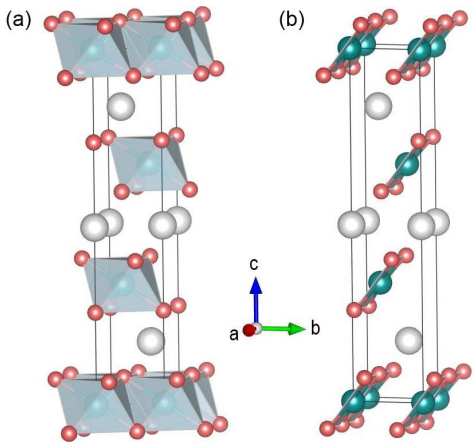}
\caption{Relaxed structures of (a) LiCoO$_{2}$ and (b) LiNiO$_{2}$. Large (gray) spheres are Li, medium (blue) spheres are Co/Ni, and small (red) spheres are O. Jahn-Teller distortion is observed in \ce{LiNiO2}; there are four short Ni$-$O bonds (shown in the figure) and two long Ni$-$O bonds (not shown). All structural figures are generated using the VESTA visualization package.\cite{VESTA}}
\label{fig;struct}
\vspace{0.3cm}
\end{figure}

Layered oxides \ce{LiMO2} were reported to crystallize in space group $R\overline{3}m$. The experimental lattice parameters are $a=2.83$ {\AA} and $c=14.12$ {\AA} (M=Co);\cite{Laubach2009} and $a=2.88$ {\AA} and $c=14.19$ {\AA} (M=Ni).\cite{Li1993} Figure~\ref{fig;struct} shows the relaxed structures of \ce{LiMO2} in hexagonal representations. The calculated lattice parameters in \ce{LiCoO2} are $a=2.80$ {\AA} and $c=14.03$ {\AA}; the Co$-$O bond length is 1.91 {\AA}. In \ce{LiNiO2}, $a=2.86$ {\AA} and $c=14.17$ {\AA}, taken as an average hexagonal unit cell; there are four short Ni$-$O bonds (1.88 {\AA}) and two long Ni$-$O bonds (2.13 {\AA}) due to the Jahn-Teller distortion associated with low-spin Ni$^{3+}$, in reasonable agreement with the experimental values of 1.91 and 2.07 {\AA} in Li$_{0.98}$Ni$_{1.02}$O$_{2}$ reported by Delmas {\it et al.}\cite{Delmas1997} The calculated magnetic moment is 0 $\mu_{\rm B}$ for Co, indicating that Co$^{3+}$ is in the low-spin state in \ce{LiCoO2}; and 0.85 $\mu_{\rm B}$ for Ni, {\it i.e.}, Ni$^{3+}$ is also in the low-spin state in \ce{LiNiO2}. The calculated formation enthalpies at 0 K are $-$6.96 eV (M=Co) and $-$6.10 eV (M=Ni), in agreement with the experimental values of $-$7.03 eV (M=Co) and $-$6.15 eV (M=Ni).~\cite{Wang2004167}

The implementation of finite-cell-size corrections in the Freysoldt approach requires values for the static dielectric constant.\cite{Freysoldt,Freysoldt11} The electronic contribution to the static dielectric constant can be obtained from the real part of the dielectric function $\epsilon_{1}(\omega)$ for $\omega\rightarrow0$. The ionic contribution, on the other hand, can be calculated using density functional perturbation theory.~\cite{Wu2005,dielectricmethod} Since the ionic contribution only depends on the Born effective charges and the vibrational modes, which are usually well described in GGA,~\cite{CZTS} this term can be calculated using GGA or GGA$+U$. We find the electronic contributions are 4.67 (for M=Co) and 5.11 (M=Ni), obtained from HSE06 calculations, whereas the ionic contributions are 8.35 (for M=Co) and 10.34 (M=Ni), obtained from GGA$+U$ calculations. The calculated total static dielectric constants are thus 13.02 for \ce{LiCoO2} and 15.45 for \ce{LiNiO2}.

\begin{figure}
\centering
%\vspace{-0.3cm}
\includegraphics[height=6.5cm]{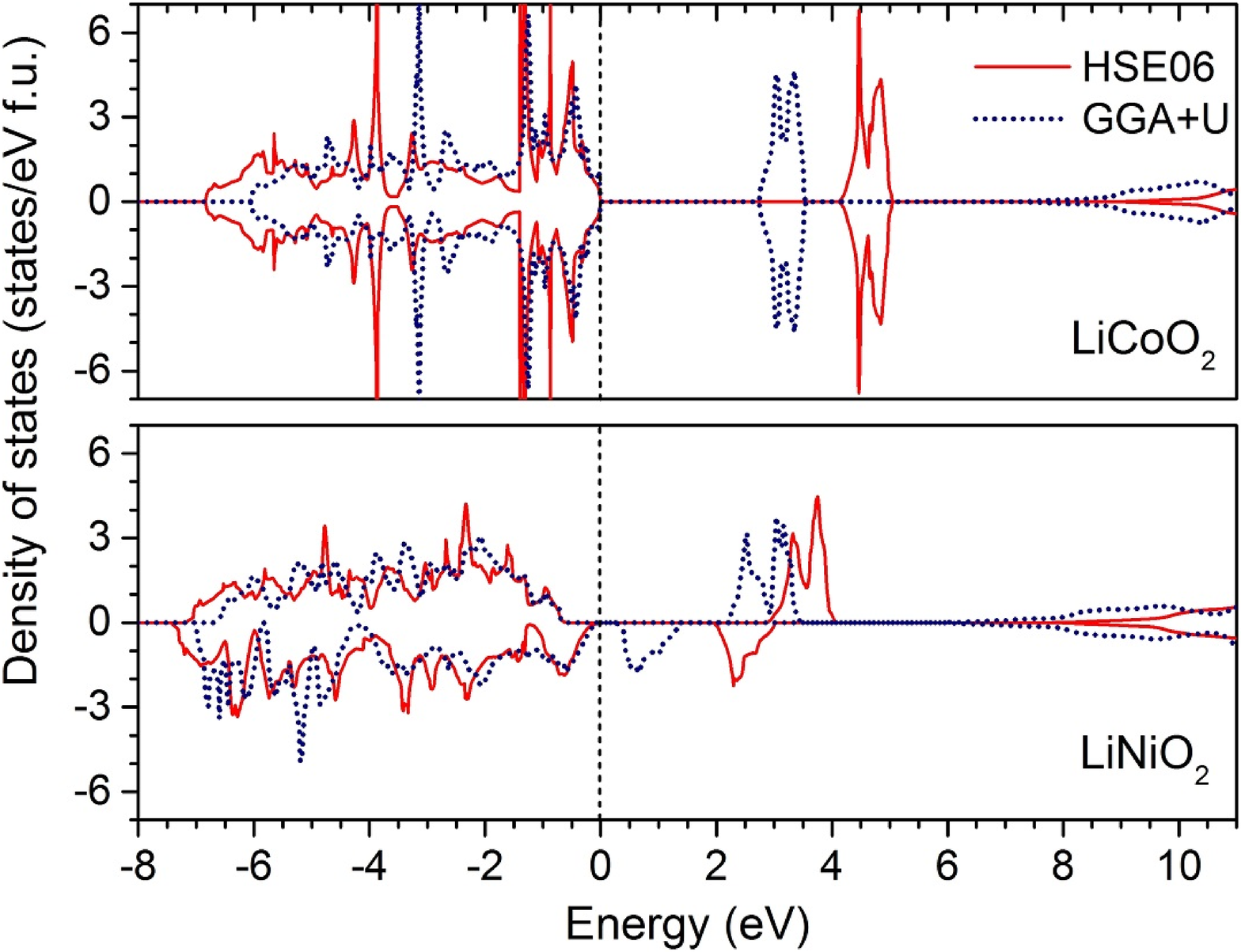}
%\vspace{-0.8cm}
\caption{Density of states of LiCoO$_{2}$ and LiNiO$_{2}$ obtained in HSE06 and GGA+$U$ calculations; $U$=4.91 eV for Co and 6.70 eV for Ni.}
\label{fig;dos}
\end{figure}

\begin{table}
\small
\caption{\ Percentage of transition-metal character at the VBM and CBM in \ce{LiMO2} from HSE06 and GGA$+U$ calculations. The character not attributable to the transition metal at the VBM and CBM comes almost exclusively from oxygen.}
\label{tbl;nature}
\begin{tabular*}{0.5\textwidth}{@{\extracolsep{\fill}}lcccc}
\hline
&\multicolumn{2}{c}{HSE06}&\multicolumn{2}{c}{GGA$+U$} \\
&VBM&CBM&VBM&CBM \\
\hline
\ce{LiCoO2}&62\%&81\%&56\%&79\% \\
\ce{LiNiO2}&34\%&65\%&24\%&48\% \\
\hline \\
%\vspace{0.05cm}
\end{tabular*}
\end{table}

Figure~\ref{fig;dos} shows the total electronic densities of states of \ce{LiMO2}, obtained in calculations using HSE06 where the default Hartree-Fock mixing parameter value $\alpha$=0.25 was used, and GGA$+U$ where $U$=4.91 eV for Co and 6.70 eV for Ni. At first glance, both methods give quite similar electronic densities of states, except that the HSE06 band gap and valence-band width are larger. The calculated band gaps are 4.11 and 1.93 eV for M=Co and Ni, respectively, in HSE06, and 2.74 eV and 0.30 eV in GGA$+U$. The HSE06 band gap for \ce{LiCoO2} is in good agreement with that of 4.2 eV reported by Ensling {\it et al.},\cite{Ensling2010} obtained in calculations using the B3LYP hybrid functional.  Upon further examination, we find that the nature of the electronic states at the valence-band maximum (VBM) and conduction-band minimum (CBM) can be dependent on the method used, especially in the case of \ce{LiNiO2}. Table~\ref{tbl;nature} lists the percentage of transition-metal character at the VBM and CBM in \ce{LiMO2} from HSE06 and GGA$+U$. The VBM and CBM in \ce{LiCoO2} are predominantly transition-metal 3$d$ states in both HSE06 and GGA$+U$. In \ce{LiNiO2}, the VBM has a significantly larger contribution from O 2$p$ states, with 34\% from the Ni atom and 32\% from each O atom (HSE06), or 24\% from the Ni atom and 38\% from each O atom (GGA$+U$). In both compounds, the Li 2$s$ state is high up in the conduction band, suggesting that Li donates one electron to the lattice and becomes Li$^{+}$. This information about the electronic structure will become very useful when we discuss defects in \ce{LiMO2} since the formation of a defect often involves removing (adding) electrons from (to) the VBM (CBM).\cite{hoang2011}

There are discrepancies in the experimental band gap values for \ce{LiCoO2} reported in the literature. Ghosh {\it et al.}\cite{Ghosh2007} obtained a band gap of 1.7 eV from optical spectroscopy. With ultraviolet-visible spectroscopy, Kushida {\it et al.}\cite{Kushida2002} found 2.1 eV. Rosolen {\it et al.}\cite{Rosolen2001253} reported a direct band gap of 2.5 eV with photocurrent spectra. Using a combination of bremsstrahlung isochromat spectroscopy and x-ray photoemission spectroscopy measurements, van Elp {\it et al.}\cite{vanElp} obtained a band gap of 2.7$\pm$0.3 eV. The discrepancies suggest that the experimental band gap value could be sensitive to the quality of the samples which in turn depends on the synthesis conditions. For \ce{LiNiO2}, Anisimov {\it et al.}\cite{anisimov1991} described the material as a small-gap insulator and cited a band gap value of 0.4 eV from inverse photoemission. Molenda {\it et al.},\cite{Molenda2002} on the other hand, reported a band gap of 0.5 eV. Apparently, our HSE06 calculations significantly overestimate the band gap values for \ce{LiMO2}. It has also been observed in some other complex oxides that HSE06 tends to overestimate the band gaps.\cite{C0JM03852K,C0CP02562C} However, it should also be noted again that no stoichiometric samples of LiNiO$_2$ exist from which to measure the gap. All known samples are defected at some level and this changes both the chemical composition and the long range structural order, as evidenced by the local, rather than cooperative, observed Jahn-Teller 
distortion.\cite{Rougier1995} This obviously complicates comparison between calculated and observed bulk properties, including the band gap. Although this issue needs further investigations from both the computational and experimental sides, it does not play a crucial role in our discussion of the energetics of point defects in \ce{LiMO2}. In fact, it has been observed that the defect formation energy at the Fermi level determined by the charge neutrality condition (\ref{eq:neutrality}) is usually not sensitive to the calculated band gap, as long as the physics near the band edges is well reproduced by the calculations.\cite{Hoang201353,Roy}

\begin{figure*}
\centering
\includegraphics[height=6.5cm]{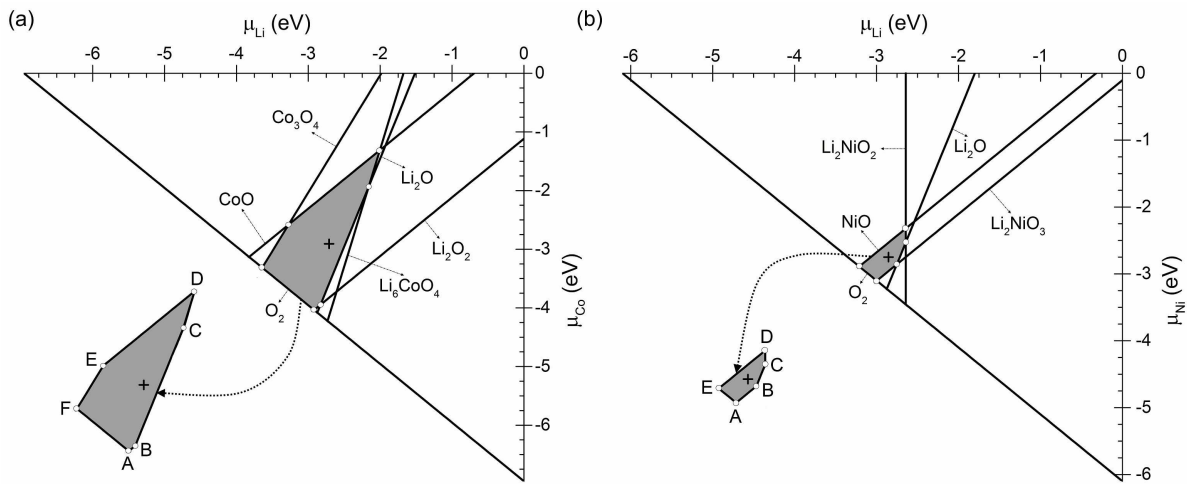}
\caption{Chemical-potential diagrams for LiMO$_{2}$: (a) M=Co and (b) M=Ni. Only phases that define the stability region of LiMO$_{2}$ are included, here shown as a shaded polygon. Point X, marked by a cross, corresponds to $\mu_{\rm{Li}}$=$-$2.71, $\mu_{\rm{Co}}$=$-$2.91, and $\mu_{\rm{O}}$=$-$0.67 eV in LiCoO$_{2}$, and $\mu_{\rm{Li}}$=$-$2.85, $\mu_{\rm{Ni}}$=$-$2.75, and $\mu_{\rm{O}}$=$-$0.25 eV in LiNiO$_{2}$. The stability region of \ce{LiCoO2} is much larger than that of \ce{LiNiO2}.}
\label{fig;chempot}
\end{figure*}

\subsection{Chemical potentials and phase stability}

\begin{table}
%\small
\caption{\ Calculated formation enthalpies at 0 K, in eV per formula unit. Experimental values at 298 K are also included}
\label{tbl;enthalpies}
\begin{tabular*}{0.5\textwidth}{@{\extracolsep{\fill}}lccc}
\hline
System & Crystal structure &This work & Experiments \\
\hline
\ce{Li2O} & cubic & $-$5.75 & $-$6.21\textit{$^{a}$} \\ %, $-$6.20\textit{$^{b}$} \\
\ce{Li2O2} & hexagonal & $-$5.84 &$-$6.56\textit{$^{a}$} \\
\ce{CoO} & hexagonal & $-$3.13 & $-$2.46\textit{$^{a}$} \\ %, $-$2.47\textit{$^{b}$} \\
\ce{Co3O4}& cubic & $-$9.94 & $-$9.43\textit{$^{a}$} \\
\ce{LiCoO2} & trigonal & $-$6.96 & $-$7.03\textit{$^{b}$}\\
\ce{Li6CoO4} & tetragonal & $-$20.62 &  \\
\ce{NiO} & cubic & $-$2.89 & $-$2.48\textit{$^{b}$} \\
\ce{LiNiO2} & monoclinic & $-$6.10 & $-$6.15\textit{$^{b}$}\\
\ce{Li2NiO2} & trigonal & $-$8.76 & \\
\ce{Li2NiO3} & monoclinic & $-$9.10 &\\
\hline \\
%\vspace{0.05cm}
\end{tabular*}
\textit{$^{a}$}~Ref.~\cite{chase}. \textit{$^{b}$}~Ref.~\cite{Wang2004167}.
\end{table}

Figure~\ref{fig;chempot} shows the atomic chemical-potential diagrams associated with \ce{LiMO2}. In order to construct these diagrams, we explored and calculated all possible Li$-$M$-$O phases available in the Materials Project database.\cite{Jain2013} As mentioned in Sec.~2.2, the zero reference state of the oxygen chemical potential $\mu_{\mathrm{O}}$ is chosen to be half of the total energy of an isolated O$_{2}$ molecule. In our calculations, the O$-$O bond in an O$_{2}$ molecule is 1.21 {\AA}, and the calculated binding energy with respect to spin-polarized O atoms is 5.16 eV, in excellent agreement with the experimental binding energy of 5.12 eV.\cite{chase} The range of Li, M, and O chemical potential values in which the host materials \ce{LiMO2} are thermodynamically stable, {\it i.e.}, the shaded regions in Fig.~\ref{fig;chempot}, are defined by the competing Li$-$M$-$O phases that can be in thermodynamic equilibrium with \ce{LiMO2}. The calculated formation enthalpies (at 0 K) of these phases and those of \ce{LiMO2} are listed in Table~\ref{tbl;enthalpies}.

Each point in the diagrams in Fig.~\ref{fig;chempot} corresponds to a specific set of Li, M, and O chemical potential values. Points A$-$F in Fig.~\ref{fig;chempot}(a) and points A$-$E in Fig.~\ref{fig;chempot}(b) represent limiting cases where the host materials \ce{LiMO2} are thermodynamically stable and in equilibrium with different competing phases. For example, point A in Fig.~\ref{fig;chempot}(a) is where \ce{O2}, \ce{Li2O2}, and \ce{LiCoO2} are in equilibrium; point B is where \ce{Li2O2}, \ce{Li2O}, and \ce{LiCoO2} are in equilibrium. These two limiting cases can be regarded as representing Li-excess (Co-deficient) environments. The environments at points A, B, and F in Fig.~\ref{fig;chempot}(a) and points A and E in Fig.~\ref{fig;chempot}(b) can also be considered as highly oxidizing, given the very high oxygen chemical potential.

As can be seen from Fig.~\ref{fig;chempot}, the stability region of \ce{LiCoO2} is much larger than that of \ce{LiNiO2}. For example, the oxygen chemical potential $\mu_{\rm O}$ goes from $-$1.82 eV to 0 eV in \ce{LiCoO2}, whereas in \ce{LiNiO2} it goes from $-$0.56 eV to 0 eV. We note that $\mu_{\rm O}$=$-$1.82 eV corresponds to O$_{2}$ gas at 1200$^\circ$C and 0.2 atm; $\mu_{\rm O}$=$-$0.56 eV corresponds to O$_{2}$ gas at 250$^\circ$C and 0.2 atm.\cite{chase} Our results thus indicate that stoichiometric \ce{LiCoO2} can be stable at much higher temperatures than stoichiometric \ce{LiNiO2}. We note that \ce{LiNiO2} would be unstable toward competing Li$-$Ni$-$O phases if the Jahn-Teller distortion were not allowed. The rhombohedral-to-monoclinic distortion lowers the total energy and hence the formation enthalpy of \ce{LiNiO2} by 0.42 eV per formula unit, stabilizing the monoclinic phase.

\subsection{Defect structure and energetics}

We investigated various intrinsic point defects in \ce{LiMO2} in all possible charge states. These defects include hole and electron polarons, hereafter denoted as $\eta^{+}$ and $\eta^{-}$; lithium vacancies ($V_{\rm Li}$) and interstitials (Li$_{i}$); lithium antisites (Li$_{\rm M}$); transition-metal antisites (M$_{\rm Li}$) and vacancies ($V_{\rm M}$); oxygen vacancies ($V_{\rm O}$); and MO$_{2}$ vacancies ($V_{{\rm MO}_{2}}$). We also considered defect complexes such as lithium divacancies ($DV_{\rm{Li}}$), antisite defect pairs (M$_{\rm{Li}}$$-$Li$_{\rm{M}}$), and a complex of M$_{\rm{Li}}$ and $V_{\rm{Li}}$.

\begin{figure*}
\centering
%\vspace{-1cm}
%\hspace{-.5cm}
\includegraphics[height=7.5cm]{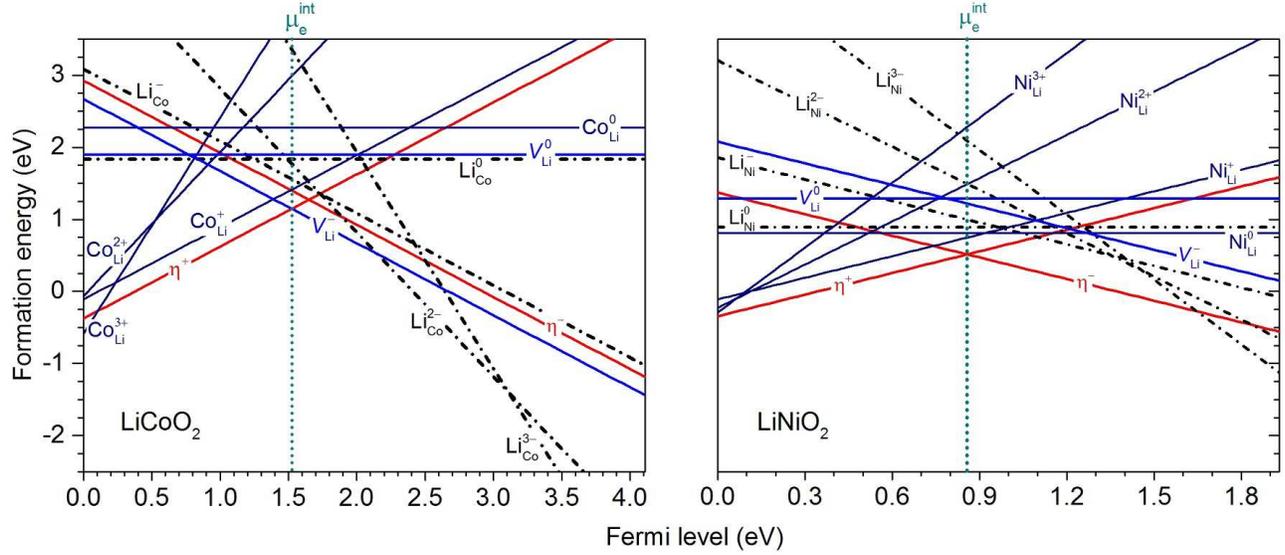}
%\vspace{0cm}
\caption{Calculated formation energies of intrinsic point defects in LiCoO$_{2}$ and LiNiO$_{2}$, plotted as a function of the Fermi level with respect to the VBM. The energies are obtained at point X, marked by a cross in the chemical-potential diagrams in Figs.~\ref{fig;chempot}(a) and \ref{fig;chempot}(b). In the absence of extrinsic charged impurities, the Fermi level of the system is at $\mu_{e}=\mu_{e}^{\rm{int}}$, where charge neutrality is maintained.}
\label{fig;formenergy}
\end{figure*}

\begin{table*}
\caption{\ Calculated formation energies ($E^{f}$) and migration barriers ($E_{m}$) of the most relevant point defects in \ce{LiMO2}. The formation energies are obtained at points A$-$F and X in Fig.~\ref{fig;chempot}(a) for M=Co and points A$-$E and X in Fig.~\ref{fig;chempot}(b) for M=Ni. The lowest energy values associated with each set of the atomic chemical potentials are underlined. The binding energies ($E_{b}$) of defect complexes are listed in the last column.}\label{tab:formenergy}
%\begin{center}
\begin{tabular*}{\textwidth}{@{\extracolsep{\fill}}cccccccccccr}
\hline
&Defect&\multicolumn{7}{c}{$E^{f}$ (eV)}&  $E_{m}$ (eV)&Constituents&$E_{b}$ (eV) \\
&&A&B&C&D&E&F&X&&& \\
\hline
\ce{LiCoO2}
&$\eta^{+}$&\underline{0.89}&\underline{0.89}&1.35&1.59&1.18&0.81&\underline{1.15}&0.10&\\ 
&$\eta^{-}$ &1.65&1.65&\underline{1.20}&\underline{0.96}&1.37&1.74&1.40&0.32&\\ 
&$V_{\rm{Li}}^-$&1.19&1.28&1.49&1.42&\underline{0.55}&\underline{0.55}&\underline{1.15}&0.70&\\ 
&$V_{\rm{Li}}^0$&1.69&1.78&2.45&2.60&1.33&0.97&1.90&&$V_{\rm{Li}}^- + \eta^{+}$&0.40\\
&Co$_{\rm{Li}}^+$&2.08&2.08&\underline{1.20}&\underline{0.96}&\underline{0.55}&\underline{0.55}&1.41&&\\ 
&Co$_{\rm{Li}}^0$&3.19&3.19&1.85&1.38&1.38&1.75&2.27&&Co$_{\rm{Li}}^+ + \eta^{-}$&0.54\\
&Co$_{\rm{Li}}^{2+}$&3.40&3.40&2.97&2.97&2.15&1.79&2.98&&\\ 
&Li$_{\rm{Co}}^{2-}$&1.36&1.36&1.79&1.79&2.60&2.97&1.77&&\\ 
&Li$_{\rm{Co}}^-$&\underline{0.89}&\underline{0.89}&1.78&2.01&2.42&2.42&1.56&&Li$_{\rm{Co}}^{2-} + \eta^{+}$&1.36\\
&Li$_{\rm{Co}}^0$&0.92&0.92&2.26&2.73&2.73&2.36&1.84&&Li$_{\rm{Co}}^{2-} + 2\eta^{+}$&2.23\\
&Co$_{\rm{Li}}^{+}$-$V_{\rm{Li}}^{-}$&2.76&2.84&2.17&1.86&0.59&0.59&2.05&&Co$_{\rm{Li}}^{+} + V_{\rm{Li}}^{-}$&0.51\\
&Co$_{\rm{Li}}$-Li$_{\rm{Co}}$&2.34&2.34&2.34&2.34&2.34&2.34&2.34&&Co$_{\rm{Li}}^{+} + \rm{Li}_{\rm{Co}}^{2-} + \eta^{+}$&1.99 \\ 
\hline
\ce{LiNiO2}
&$\eta^{+}$&\underline{0.51}&\underline{0.51}&\underline{0.51}&\underline{0.51}&\underline{0.51}&&\underline{0.51}&0.28, 0.21&\\ 
&$\eta^{-}$&\underline{0.51}&\underline{0.51}&\underline{0.51}&\underline{0.51}&\underline{0.51}&&\underline{0.51}&0.26, 0.28&\\ 
&$V_{\rm{Li}}^-$&1.08&1.32&1.42&1.42&0.87&&1.22&0.56, 0.66&\\ 
&$V_{\rm{Li}}^0$&1.15&1.39&1.49&1.49&0.93&&1.29&&$V_{\rm{Li}}^- + \eta^{+}$&0.45\\
&Ni$_{\rm{Li}}^+$&0.96&0.96&0.76&0.53&0.53&&0.75&&\\ 
&Ni$_{\rm{Li}}^0$&1.01&1.01&0.82&0.59&0.59&&0.80&&Ni$_{\rm{Li}}^{+} + \eta^{-}$&0.45\\
&Ni$_{\rm{Li}}^{2+}$&1.70&1.70&1.50&1.27&1.27&&1.49&&Ni$_{\rm{Li}}^{+} + \eta^{+}$&$-$0.23\\
&Li$_{\rm{Ni}}^{2-}$&1.29&1.29&1.48&1.71&1.71&&1.49&&\\ 
&Li$_{\rm{Ni}}^-$&0.79&0.79&0.98&1.22&1.21&&1.00&&Li$_{\rm{Ni}}^{2-}$ + $\eta^{+}$&1.01\\
&Li$_{\rm{Ni}}^0$&0.68&0.68&0.88&1.11&1.11&&0.89&&Li$_{\rm{Ni}}^{2-}$ + 2$\eta^{+}$&1.63\\
&Ni$_{\rm{Li}}^+$-$V_{\rm{Li}}^-$&1.87&2.11&2.02&1.79&1.23&&1.80&&Ni$_{\rm{Li}}^+ + V_{\rm{Li}}^-$&0.17\\
&Ni$_{\rm{Li}}$-Li$_{\rm{Ni}}$&1.19&1.19&1.19&1.19&1.19&&1.19&&Ni$_{\rm{Li}}^+ + {\rm Li}_{\rm{Ni}}^{2-} + \eta^{+}$&1.57 \\ 
\hline
\end{tabular*}
%\end{center}
\end{table*}

Figure~\ref{fig;formenergy} shows the calculated formation energies of selected point defects in \ce{LiMO2}, obtained at point X in the thermodynamically allowed regions of the chemical potential diagrams, marked by a cross in Figs.~\ref{fig;chempot}(a) and \ref{fig;chempot}(b). The slope in the formation energy plots indicates the charge state. Positively charged defects have positive slopes; negatively charged defects have negative slopes. We find that charged defects have positive formation energies only near midgap. Therefore, any attempt to deliberately shift the Fermi level to the VBM or CBM, {\it e.g.}, via doping with acceptors or donors, will result in positively or negatively charged intrinsic defects having negative formation energies, {\it i.e.}, the intrinsic defects will form spontaneously and counteract the effects of doping. This indicates that intrinsic point defects in \ce{LiMO2} cannot act as sources of band-like electrons and holes, and the material cannot be made $n$-type or $p$-type. In the absence of electrically active impurities that can shift the Fermi-level position or when such impurities occur in much lower concentrations than charged intrinsic defects, the Fermi level $\mu_{e}$ is determined by the charge neutrality condition (\ref{eq:neutrality}), hereafter this position is denoted as $\mu_{e}^{\rm int}$, and is almost exclusively defined by the positively and negatively charged point defects with the lowest formation energies.\cite{hoang2009,wilson-short,hoang2011,amidePRB} With the chosen sets of atomic chemical potentials, $\mu_{e}^{\rm int}$ is 1.52 eV above the VBM in \ce{LiCoO2}, determined by the hole polaron $\eta^{+}$ and the negatively charged lithium vacancy $V_{\rm Li}^{-}$, or 0.88 eV above the VBM in \ce{LiNiO2}, determined by $\eta^{+}$ and the electron polaron $\eta^{-}$.

The results presented in Fig.~\ref{fig;formenergy} are, however, not the only scenario that may occur in \ce{LiMO2}. As it is clear from eqn.~(\ref{eq:eform}), the defect formation energy and hence concentration depend on the chemical potentials. We therefore list in Table~\ref{tab:formenergy} the calculated formation energies of the most relevant point defects in \ce{LiMO2} for different sets of atomic chemical potentials which correspond to points A$-$F and X in Fig.~\ref{fig;chempot}(a) for M=Co and points A$-$E and X in Fig.~\ref{fig;chempot}(b) for M=Ni, obtained at the Fermi-level position $\mu_{e}^{\rm int}$ determined by the charge neutrality condition. With the allowed ranges of the atomic chemical potentials, we find that $\mu_{e}^{\rm int}$ is in the range of 1.19$-$1.96 eV in \ce{LiCoO2}, which is always far away from both the VBM and CBM. In \ce{LiNiO2}, $\mu_{e}^{\rm int}$ is always at 0.88 eV. The results summarized in Table~\ref{tab:formenergy} clearly show that the point defect landscapes in \ce{LiMO2} can be very different under different thermodynamic conditions. In the following, we analyze in detail the structure and energetics of the defects.

{\bf Small polarons}. The formation of $\eta^{+}$ involves removing an electron from the system which results in a low-spin M$^{4+}$ ion at the M$^{3+}$ lattice site. The calculated magnetic moment of the M$^{4+}$ ion is 1.13 $\mu_{\rm B}$ for M=Co or 0.07 $\mu_{\rm B}$ for M=Ni. The local lattice geometry near the M$^{4+}$ ion is slightly distorted with the six neighboring O atoms moving toward M$^{4+}$. In \ce{LiCoO2}, the Co$-$O bond length at the Co$^{4+}$ site is 1.88 {\AA}, as compared to 1.91 {\AA} of the other Co$-$O bonds. In \ce{LiNiO2}, there are four Ni$-$O bonds of 1.86 {\AA}, as compared to 1.88 {\AA} of the other four short Ni$-$O bonds, and two Ni$-$O bonds of 1.90 {\AA}, as compared to 2.13 {\AA} of the other long Ni$-$O bonds, {\it i.e.}, the Jahn-Teller distortion almost completely vanishes at the Ni$^{4+}$ site. The creation of $\eta^{-}$, on the other hand, corresponds to adding an electron to the system which results in a high-spin M$^{2+}$ ion at the M$^{3+}$ lattice site. The magnetic moment of the M$^{2+}$ ion is 2.66 $\mu_{\rm B}$ for M=Co or 1.66 $\mu_{\rm B}$ for M=Ni. The high-spin state of M$^{2+}$ is lower in energy than the low-spin state by 0.54 eV in the case of M=Co. We note that in bulk \ce{LiMO2}, the M$^{3+}$ ion is most stable in its low-spin state. The local geometry near the M$^{2+}$ ion is also slightly distorted, but with the neighboring O atoms moving away from M$^{2+}$; the six Co$-$O bonds at the Co$^{2+}$ site in \ce{LiCoO2} are now 2.03 {\AA} and the Ni$-$O bonds at the Ni$^{2+}$ site in \ce{LiNiO2} are now 2.00 {\AA} (four bonds) and 2.12 {\AA} (two bonds). The calculated formation energies of $\eta^{+}$ and $\eta^{-}$ in \ce{LiCoO2} are in the ranges of 0.81$-$1.59 eV and 0.96$-$1.74eV, respectively, depending on the specific set of the atomic chemical potentials. In \ce{LiNiO2}, the formation energy of $\eta^{+}$ and $\eta^{-}$ is always 0.51 eV, independent of the atomic chemical potentials.

\begin{figure}
%\vspace{0.2cm}
\centering
\includegraphics[height=6cm]{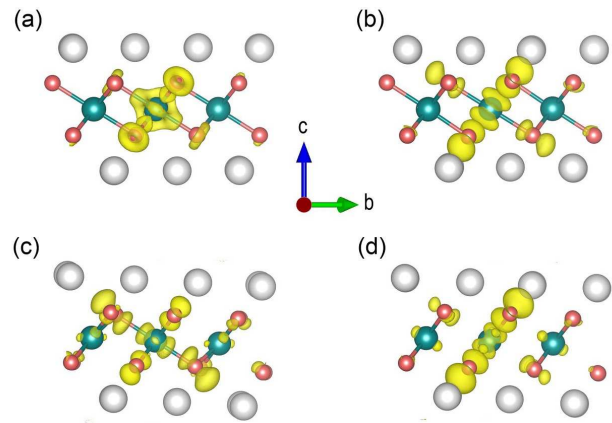}
\caption{Charge density of the small hole and electron polarons in layered oxides: (a) $\eta^{+}$ and (b) $\eta^{-}$ in \ce{LiCoO2}, and (c) $\eta^{+}$ and (d) $\eta^{-}$ in \ce{LiNiO2}. Large (gray) spheres are Li, medium (blue) spheres are Co/Ni, and small (red) spheres are O.}
\label{fig;polarons}
\end{figure}

In forming the polarons, the removed or added electron can be regarded as becoming self-trapped in the local lattice distortion, which acts as a potential well, induced by its own presence. Since the distortion is found to be mainly limited to the neighboring O atoms of the M$^{4+}$ or M$^{2+}$ ion, these hole and electron polarons can be considered as small polarons.\cite{Stoneham2007,Shluger1993} Figure~\ref{fig;polarons} shows the charge density of the polarons in \ce{LiMO2}. Most of the positive (hole) or negative (electron) charge resides on the transition metal, but significant charge is also on surrounding oxygens, particularly in the case of \ce{LiNiO2}. The features of $\eta^{+}$ are necessarily related to the nature of the VBM from which the electrons are removed to form the polarons, and those of $\eta^{-}$ are related to the nature of the CBM to which electrons are added. The stability of a polaron in a given material can be assessed through its self-trapping energy,\cite{Shluger1993,Varley2012,JACE:JACE4771} defined as the difference between the formation energy of the free hole or electron and that of the hole or electron polaron. In \ce{LiCoO2}, we find the self-trapping energies of $\eta^{+}$ and $\eta^{-}$ are 0.36 and 1.35 eV, respectively; in \ce{LiNiO2}, the self-trapping energies are 0.36 and 0.65 eV for $\eta^{+}$ and $\eta^{-}$. Our HSE06 calculations using a smaller-than-default Hartree-Fock mixing parameter, particularly, $\alpha$=0.15, also show that the polarons are stable, though with smaller self-trapping energies: 0.12 and 0.62 eV for $\eta^{+}$ and $\eta^{-}$ in \ce{LiCoO2}, and 0.16 and 0.28 eV for $\eta^{+}$ and $\eta^{-}$ in \ce{LiNiO2}. GGA$+U$ calculations with $U$=6.70 eV, on the other hand, cannot stabilize a hole polaron in \ce{LiNiO2}. This is because the VBM in GGA$+U$ is predominantly delocalized O 2$p$ states: the Ni atom contributes only 24\% whereas each O atom contributes 38\%; see Table~\ref{tbl;nature}. We note that Koyama {\it et al.}~also found in GGA$+U$ calculations with $U$=5 eV that the hole in \ce{LiNiO2} is delocalized.\cite{Koyama2012} 

{\bf Vacancies and interstitials}. The formation of $V_{\rm Li}^-$ involves removing a Li$^{+}$ ion, which causes inward movement of the four neighboring Li atoms by 0.11 {\AA} toward the void formed by the removed Li$^{+}$. The calculated formation energy of $V_{\rm Li}^-$ can be as low as 0.55 eV and as high as 1.49 eV in \ce{LiCoO2}, depending on the atomic chemical potentials; in \ce{LiNiO2}, it is in the range of 0.87$-$1.42 eV. $V_{\rm Li}^0$ is created by removing a Li atom, which is in fact a Li$^{+}$ ion and an electron from a neighboring M atom, leading to the formation of a void at the site of the removed Li$^{+}$ and an M$^{4+}$ at the neighboring M site. There is a slight distortion in the local geometry near the M$^{4+}$ ion and the magnetic moment at this site is 1.13 $\mu_{\rm B}$ (M=Co) or 0.05 $\mu_{\rm B}$ (M=Ni), similar to those for the hole polarons. $V_{\rm Li}^0$ is thus a complex of $V_{\rm Li}^-$ and $\eta^{+}$ with a binding energy of 0.40 eV (M=Co) or 0.45 (M=Ni). This defect structure also suggests that for each Li atom removed from \ce{LiMO2} cathodes, {\it e.g.}, during delithiation, the material is left with one negatively charged lithium vacancy and one hole polaron.

For lithium interstitials, the lowest-energy configuration is Li$_{i}^{+}$, created by adding Li$^{+}$ into the system. The addition of Li$^{+}$ to a Li layer results in a significant rearrangement of the Li$^{+}$ in that layer with at least two Li$^{+}$ moving away from their original octahedral sites and toward the tetrahedral sites. Because of this large rearrangement, Li$_{i}^{+}$ has a rather high formation energy, 2.25$-$3.19 eV (M=Co) or 1.44$-$1.99 eV (M=Ni). With that high formation energy, lithium interstitials are not likely to form in \ce{LiCoO2}, which is consistent with experiments where no extra Li was found at the tetrahedral sites even in lithium over-stoichiometric Li$_{1+x}$CoO$_{2}$ ($x>0$).\cite{Levasseur2000}

Other vacancies include $V_{\rm O}$, $V_{\rm M}$, and $V_{{\rm MO}_{2}}$, created by removing O, M, and MO$_{2}$ units, respectively. The oxygen vacancies can be stable as $V_{\rm O}^0$, $V_{\rm O}^+$, and $V_{\rm O}^{2+}$ whose formation energies are 2.08 eV or higher in \ce{LiCoO2} and 1.43 eV or higher in \ce{LiNiO2}. We find that some neighboring Co$^{3+}$ ions of the oxygen vacancy in \ce{LiCoO2} possess an intermediate-spin (IS) state with a calculated magnetic moment of 1.73 $\mu_{\rm B}$, as compared to 0 $\mu_{\rm B}$ of low-spin Co$^{3+}$ and 3.12 $\mu_{\rm B}$ of high-spin Co$^{3+}$. However, with the high calculated formation energy in \ce{LiCoO2}, oxygen vacancies (and the associated IS Co$^{3+}$) are unlikely to form. This may explain why there has been no experimental evidence for the presence of oxygen vacancies in \ce{LiCoO2}.\cite{Hertz2008} We note that the IS Co$^{3+}$ has been proposed to be present in associated with oxygen vacancies in Li-excess Li$_{1+x}$CoO$_{2}$ by Levasseur {\it et al.}\cite{Levasseur2003} to explain for the observed paramagnetism. Several authors have also reported in their theoretical works that the IS state can be stabilized either in the bulk\cite{Koyama2012,Carlier2013} or at the surface;\cite{QianIS2012} however, these authors did not comment on the energetics of the oxygen vacancy. Regarding the transition-metal vacancies, $V_{\rm M}$ can be stable as $V_{\rm M}^{0}$, $V_{\rm M}^{-}$, $V_{\rm M}^{2-}$, or $V_{\rm M}^{3-}$ with calculated formation energies of 2.78 eV or higher in \ce{LiCoO2} and 2.63 eV or higher in \ce{LiNiO2}. Finally, $V_{{\rm MO}_{2}}$ can be stable as $V_{{\rm MO}_{2}}^0$ or $V_{{\rm MO}_{2}}^+$ whose formation energies are found to be 4.98 eV or higher in \ce{LiCoO2} and 2.96 eV or higher in \ce{LiNiO2}. $V_{{\rm MO}_{2}}^0$ is in fact a complex of $V_{{\rm MO}_{2}}^+$ and $\eta^{-}$ with a binding energy of 1.88 eV (M=Co) or 0.37 eV (M=Ni). These vacancies all have very high calculated formation energies and are therefore not included in Fig.~\ref{fig;formenergy} and Table~\ref{tab:formenergy}.

{\bf Antisite defects}. Lithium antisites Li$_{\rm M}$ are created by replacing M at an M site with Li. Li$_{\rm M}^{2-}$ is Li$^{+}$ replacing M$^{3+}$. Due to the Coulombic interaction, the six nearest Li$^{+}$ ion neighbors of Li$_{\rm M}^{2-}$ are pulled closer to the negatively charged defect with the Li$_{\rm Co}^{2-}$$-$Li distance being 2.67 {\AA}, compared to 2.85 {\AA} of the equivalent Co$-$Li distance in bulk \ce{LiCoO2}; in \ce{LiNiO2} the average Li$_{\rm Ni}^{2-}$$-$Li distance is 2.72 {\AA}, compared with 2.89 {\AA} of the equivalent Ni$-$Li distance in the bulk. Li$_{\rm M}^{-}$, on the other hand, can be regarded as a complex of Li$_{\rm M}^{2-}$ and $\eta^{+}$ with the distance between the two defects being 2.76 {\AA} (M=Co) or 2.92 {\AA} (M=Ni). The binding energy of Li$_{\rm M}^{-}$ with respect to Li$_{\rm M}^{2-}$ and $\eta^{+}$ is 1.36 eV (M=Co) or 1.01 eV (M=Ni). Similarly, Li$_{\rm M}^{0}$ is a complex of Li$_{\rm M}^{2-}$ and two $\eta^{+}$, with the binding energy being 2.23 eV (M=Co) or 1.63 eV (M=Ni). Among the lithium antisites, the calculated formation energy of Li$_{\rm Co}^{-}$ in \ce{LiCoO2} can be as low as 0.89 eV and that of Li$_{\rm Ni}^{0}$ in \ce{LiNiO2} can be as low as 0.68 eV, the values obtained at points A and B in the chemical potential diagrams in Figs.~\ref{fig;chempot}(a) and \ref{fig;chempot}(b).

Transition-metal antisites M$_{\rm Li}$ are created in a similar way by replacing Li at a Li site with M. We find that in M$_{\rm Li}^{+}$ the transition metal is stable as high-spin M$^{2+}$ with the calculated magnetic moment of 2.69 $\mu_{\rm B}$ (M=Co) or 1.70 $\mu_{\rm B}$ (M=Ni). The calculated formation energy of Co$_{\rm Li}^{+}$ is as low as 0.55 eV in \ce{LiCoO2} and that of Ni$_{\rm Li}^{+}$ is as low as 0.53 eV in \ce{LiNiO2}; see Table~\ref{tab:formenergy}. For M$_{\rm Li}^{2+}$, we find that Co$_{\rm Li}^{2+}$ can be regarded as replacing Li$^{+}$ with high-spin Co$^{3+}$ which has a calculated magnetic moment of 3.13 $\mu_{\rm B}$. This defect has the formation energy in the range of 1.79$-$3.40 eV, which is much higher than that of Co$_{\rm Li}^{+}$. Ni$_{\rm Li}^{2+}$, on the other hand, consists of Ni$_{\rm Li}^{+}$ and $\eta^{+}$, but these two defects are not stable as a unit because of the repulsive Coulomb interaction; the binding energy Ni$_{\rm Li}^{2+}$ with respect to Ni$_{\rm Li}^{+}$ and $\eta^{+}$ is $-$0.23 eV. Finally, M$_{\rm Li}^{0}$ can be regarded as a complex of M$_{\rm Li}^{+}$ and $\eta^{-}$ with a binding energy of 0.54 eV (M=Co) or 0.45 eV (M=Ni). We find the energy of high-spin Co$_{\rm Li}$ is significantly lower than that of metastable, low-spin Co$_{\rm Li}$, by 1.23, 0.92, 0.25, or 0.29 eV when charge state of the defect is 0, 1+, 2+, or 3+, respectively. 

{\bf Defect complexes}. The defects presented above can be categorized into elementary intrinsic defects ({\it e.g.}, $\eta^{+}$, $\eta^{-}$, $V_{\rm Li}^{-}$, M$_{\rm Li}^{+}$, and Li$_{\rm M}^{2-}$) and defect complexes ({\it e.g.}, $V_{\rm Li}^{0}$) whose structure and energetics can be interpreted in terms of those of the former. In addition to the polaron-containing complexes, we also considered lithium divacancies. $DV_{\rm Li}^{2-}$ consists of two $V_{\rm Li}^-$ on the nearest-neighboring sites. This defect has a calculated formation energy of 1.65$-$3.53 eV in \ce{LiCoO2} or 4.79$-$5.91 eV in \ce{LiNiO2}. In both compounds, it has a negative binding energy of $-$0.55 (M=Co) or $-0.43$ eV (M=Ni), indicating that at the lithium concentration in our calculations the divacancy is unstable toward its individual constituents. $DV_{\rm Li}^{0}$ is a complex of two $V_{\rm Li}^-$ and two $\eta^{+}$ with a binding of 0.99 eV (M=Co) or 1.13 eV (M=Ni) with respect to its individual constituents. Its calculated formation energy is in the range of 1.73$-$5.01 eV (M=Co) or 1.62$-$2.74 eV (M=Ni). Other defect complexes include M$_{\rm{Li}}^{+}$$-$Li$_{\rm{M}}^{-}$ and M$_{\rm{Li}}^{+}$$-$$V_{\rm{Li}}^{-}$. The antisite defect pair is a complex of M$_{\rm{Li}}^{+}$, Li$_{\rm{M}}^{2-}$, and $\eta^{+}$ with a binding energy of 1.99 eV (M=Co) or 1.57 eV (M=Ni), and has a formation energy of 2.34 eV (M=Co) or 1.19 eV (M=Ni). M$_{\rm{Li}}^{+}$$-$$V_{\rm{Li}}^{-}$, on the other hand, has a binding energy of 0.51 (M=Co) or 0.17 eV (M=Ni) and a formation energy as low as 0.59 eV (M=Co) or 1.23 (M=Ni). The structure and energetics of some of these complexes are summarized in Table~\ref{tab:formenergy}.

We note that a defect complex is not necessarily stable as a single unit, even if it has a finite, positive binding energy and a low formation energy; as discussed in Ref.~\cite{walle:3851}, if the binding energy is smaller than the formation energy of the constituents, entropy favors the formation of individual defects.   

Koyama {\it et al.}\cite{Koyama2012,Koyama2013} also reported the calculated formation energies of intrinsic defects in \ce{LiMO2}. In their GGA+$U$ calculations, $U$=5 eV was used for both Co and Ni in all compounds, and corrections for finite-supercell-size effects were not included. Their results, assuming equilibrium with Li$_{2}$O and O$_{2}$ gas at 627$^\circ$C and 0.2 atm, appear to indicate that hole and electron polarons are the dominant defects in \ce{LiCoO2} with a formation energy of 0.68 eV, whereas Ni$_{\rm Li}^{0}$ is the dominant defect in \ce{LiNiO2} with a formation energy of about 0.04 eV; see Figs.~S2(a) and S2(b) in the Supporting Information of Ref.~\cite{Koyama2012}. Given the same equilibrium assumption, which translates into a set of the atomic chemical potentials that actually corresponds a point between points B and C along the Li$_{2}$O line in Fig.~\ref{fig;chempot}(a), our calculations show that $\eta^{+}$ and $\eta^{-}$ are the dominant defects in \ce{LiCoO2} but with a calculated formation energy of 1.27 eV. In \ce{LiNiO2}, the mentioned assumption translates into a set of atomic chemical potentials corresponding to a point on the Li$_{2}$O line in Fig.~\ref{fig;chempot}(b) but much higher than point C and well beyond the region where \ce{LiNiO2} is stable. In other words, the assumption that \ce{LiNiO2} is in equilibrium with Li$_{2}$O and O$_{2}$ gas at 627$^\circ$C and 0.2 atm cannot be realized. This explains why the results for \ce{LiNiO2} reported by Koyama {\it et al.}~are qualitatively different from ours.

\subsection{Defect migration}

Migration of selected intrinsic point defects in \ce{LiMO2} were investigated. For the electronic defects, the migration of a small polaron between two positions $q_{\rm A}$ and $q_{\rm B}$ can be described by the transfer of the lattice distortion over a one-dimensional Born-Oppenheimer surface.\cite{Iordanova122,Iordanova123,Maxisch:2006p103} To estimate the energy barrier, we computed the energies of a set of cell configurations linearly interpolated between $q_{\rm A}$ and $q_{\rm B}$ and identify the energy maximum. For the ionic defects, the NEB method\cite{ci-neb} was used to estimate the migration barrier for the lithium vacancy via a monovacancy or divacancy mechanism. In the monovacancy mechanism, the migration path of the isolated $V_{\rm Li}^{-}$ is calculated by moving a Li$^{+}$ ion from a nearby lattice site into the vacancy. In the divacancy mechanism, the defect structure $DV_{\rm Li}^{2-}$ was used and the migration path of one of the two lithium vacancies in the defect complex is calculated by moving a Li$^{+}$ ion from a nearby lattice site into the vacancy.

\begin{figure}
%\vspace{-0.4cm}
\centering
\hspace{-0.3cm}
\includegraphics[height=6.5cm]{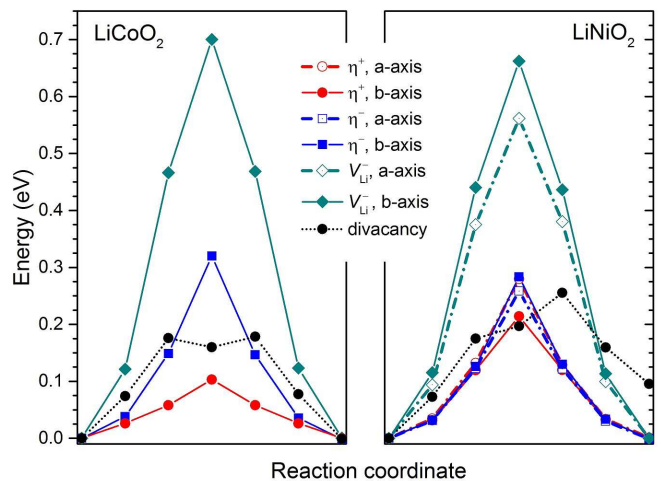}
%\vspace{-0.67cm}
\caption{Calculated migration barriers of small polarons and lithium vacancies (via monovacancy and divacancy mechanisms) in LiCoO$_{2}$ and LiNiO$_{2}$. In LiNiO$_{2}$, the migration paths along $a$- and $b$-axis are slightly different and the migration path of the divacancy is not symmetric because of the Jahn-Teller distortion.}
\label{fig;migration}
\end{figure}

Figure~\ref{fig;migration} shows the calculated migration barriers for the small hole and electron polarons and lithium vacancies in \ce{LiMO2}. We find that the migration barriers of the polarons are low: 0.10 and 0.32 eV for the hole and electron polarons in \ce{LiCoO2}, respectively, and as low as 0.21 and 0.26 eV for the hole and electron polarons in \ce{LiNiO2}; see also Table~\ref{tab:formenergy}. For the vacancies, the monovacancy mechanism gives rather high barriers: 0.70 eV in \ce{LiCoO2} and as low as 0.56 eV in \ce{LiNiO2}. The divacancy mechanism, however, gives much lower barriers: 0.18 and 0.26 eV for the lithium vacancies in \ce{LiCoO2} and \ce{LiNiO2}, respectively. We note that the migration paths along $a$- and $b$-axis in \ce{LiNiO2} are slightly different and the migration path of the divacancy is not symmetric because of the Jahn-Teller distortion. Our results for the vacancies in \ce{LiCoO2} are in agreement with those of Van der Ven and Ceder where migration barriers of about 0.8 and 0.2 eV were found for the monovacancy and divacancy mechanisms, respectively.\cite{VanderVen2000}

\section{Discussion}

It emerges from our results for \ce{LiCoO2} that defect landscapes in this compound are sensitive to the atomic chemical potentials, {\it i.e.}, the experimental conditions during synthesis. At points A and B in the chemical-potential diagram, {\it cf.}~Fig.~\ref{fig;chempot}(a), which represent highly oxidizing, Li-excess (Co-deficient) environments, the dominant intrinsic point defects are $\eta^{+}$ and Li$_{\rm{Co}}^{-}$. These two defects have a relatively low calculated formation energy (0.89 eV) and can exist as Li$_{\rm{Co}}^{0}$, a defect complex of Li$_{\rm{Co}}^{2-}$ and two $\eta^{+}$. With a binding energy of 2.23 eV with respect to its individual constituents, and a formation energy of 0.92 eV, Li$_{\rm{Co}}^{0}$ is expected to be stable as a unit. At points C and D, the dominant defects are Co$_{\rm{Li}}^{+}$ and $\eta^{-}$, which have a formation energy of 1.20 eV (at C) or 0.96 eV (at D). The defect complex of Co$_{\rm{Li}}^{+}$ and $\eta^{-}$, {\it i.e.}, Co$_{\rm{Li}}^{0}$, is not expected to be stable as a unit because of its small binding energy (0.54 eV). At points E and F, the dominant defects are Co$_{\rm{Li}}^{+}$ and $V_{\rm Li}^{-}$, which have a low formation energy (0.55 eV). These two defects may form as Co$_{\rm{Li}}^{+}$$-$$V_{\rm Li}^{-}$, a defect complex of Co$_{\rm{Li}}^{+}$ and $V_{\rm Li}^{-}$ which has a formation energy of 0.59 eV and a binding energy of 0.51 eV, although the complex is not expected to be stable as a unit. At point X, the dominant defects are $\eta^{+}$ and $V_{\rm Li}^{-}$, which have a formation energy of 1.15 eV. These two defects may form as $V_{\rm Li}^{0}$, a defect complex of $\eta^{+}$ and $V_{\rm Li}^{-}$ which has a formation energy of 1.90 eV, but this complex is, again, not expected to be stable as a unit because of its small binding energy (0.40 eV). Finally, the dominant defects could also be $\eta^{+}$ and $\eta^{-}$, if the atomic chemical potentials correspond to a point between B and C along the \ce{Li2O} line in Fig.~\ref{fig;chempot}(a), as discussed in Sec.~3.3. The two defects in this case, however, have a relatively high formation energy (1.27 eV).

In the above mentioned defect scenarios, there are always low-spin Co$^{4+}$ ions (in form of $\eta^{+}$) and/or high-spin Co$^{2+}$ (in form of $\eta^{-}$) associated with the dominant defects in \ce{LiCoO2}, in addition to low-spin Co$^{3+}$ ions. Our results are thus consistent with the fact that experimental studies of the magnetic properties always reveal localized magnetic moments.\cite{Chernova2011}

Cobalt antisites have been found in \ce{LiCoO2} samples, especially those synthesized at low temperatures.\cite{Gummow1992327} However, it should be noted that defect landscapes in \ce{LiCoO2} may not have a simple dependence on the synthesis temperature. The oxygen chemical potential, which is usually controlled via controlling oxygen partial pressure and temperature, is just one of several variables that define defect formation energies and thus concentrations. Other variables include Li and Co chemical potentials, which involve the {\it actual} amount of Li and Co participating in the reaction that forms \ce{LiCoO2}. In fact, we find that defect formation energies are sensitive to the Co:Li ratio. Besides, when synthesized at lower temperatures, some processes may be kinetically hindered. Overall, in order to avoid forming Co antisites one has to move away from points C$-$F and their nearby regions in the chemical-potential diagrams. Also, because of the difficulty in controlling the amount of volatile Li in the synthesis reaction, points A and B and their nearby region most likely represent the environments where one can obtain \ce{LiCoO2} samples with good electrochemical performance. This may explain why \ce{LiCoO2} in commercial applications is often made deliberately Li-excess.\cite{Chernova2011} Our results thus provide guidelines for defect-controlled synthesis and defect characterization.

In light of the results for \ce{LiCoO2}, let us re-examine a defect model for lithium over-stoichiometric \ce{LiCoO2} which can be realized in experiments by using a reaction mixture with the Li:Co molar ratio of greater than 1, {\it i.e.}, in Li-excess (Co-deficient) environments. This case is associated with the scenario obtained at points A and B in Fig.~\ref{fig;chempot}(a) as mentioned earlier. It has been suggested by several authors that the excess Li$^+$ goes into the Co$^{3+}$ site, thus forming Li$_{\rm Co}^{2-}$, and the chemical formula of the over-stoichiometric \ce{LiCoO2} can be written as Li$_{1+\delta}$Co$_{1-\delta}$O$_{2-\delta}$, where Li$_{\rm Co}^{2-}$ is charge-compensated by oxygen vacancy $V_{\rm O}^{2+}$. The paramagnetism experimentally observed in this material is thought to be due an IS state of Co$^{3+}$ that is associated with the oxygen vacancy.\cite{Levasseur2003,Carlier2013} Our studies, however, show that the dominant intrinsic defects in this case should be Li$_{\rm Fe}^{2-}$ and $\eta^{+}$. Besides, oxygen vacancies and hence the associated IS Co$^{3+}$ are unlikely to occur in bulk \ce{LiCoO2} because the vacancy formation energy is high. We suggest that the chemical formula should be written as Li$_{1+\delta}$Co$_{1-\delta}$O$_{2}$ or, more explicitly, as [Li$^{+}$]$_{1+\delta}$[Co$^{4+}$]$_{2\delta}$[Co$^{3+}$]$_{1-3\delta}$[O$^{2-}$]$_{2}$ where each Li$_{\rm Co}^{2-}$ is associated with two $\eta^{+}$ ({\it i.e.}, Co$^{4+}$), assuming that there are no extrinsic defects (impurities) in the material. The presence of low-spin Co$^{4+}$ in our defect model is consistent with experimental data reported by Hertz {\it et al.},\cite{Hertz2008} after their results have been corrected for an error in the magnetic moment calculations.~\cite{Chernova2011} We note that the formation energy of oxygen vacancies can be lower at the surface or interface, given that the bonding environment there is less constrained than in the bulk. In that case, oxygen vacancies and the associated IS Co$^{3+}$ ions may actually exist.

Regarding \ce{LiNiO2}, there are three major observations that can be drawn from our results. Firstly, \ce{LiNiO2} is less stable than \ce{LiCoO2}, as suggested by the chemical-potential diagrams in Fig.~\ref{fig;chempot}. Stoichiometric \ce{LiNiO2} can only be stable at much lower temperatures than stoichiometric \ce{LiCoO2}. Secondly, $\eta^{+}$ and $\eta^{-}$ are always the dominant intrinsic point defects and have a relatively small formation energy (0.51 eV), independent of the atomic chemical potentials. This indicates that a certain amount of Ni$^{3+}$ ions in \ce{LiNiO2} undergo charge disproportionation: 2Ni$^{3+}$ $\rightarrow$ Ni$^{4+}$ + Ni$^{2+}$, where Ni$^{4+}$ and Ni$^{2+}$ are stable in form of $\eta^{+}$ and $\eta^{-}$. Finally, nickel antisites Ni$_{\rm Li}^{+}$ have a low formation energy, as low as 0.53 eV and only as high as 0.96 eV. This low formation energy is thus consistent with the high concentration of Ni in the Li layers as reported in experiments.\cite{Barra1999,Chappel2002,Kalyani2005,Chernova2011} These individual issues, separately or in combination, must probably be responsible for the experimental observations reported in the literature, which include the difficulties in synthesizing \ce{LiNiO2} at high temperatures, the absence of long-range Jahn-Teller distortion and magnetic ordering, and the poor electrochemical performance.\cite{Whittingham_CR,Dutta1992123,Hirano1995,Kanno1994216,Chernova2011,Kalyani2005,Barra1999,Chappel2002} Our results also suggest that tuning the synthesis conditions would not remove the charge disproportionation from \ce{LiNiO2}. The concentration of nickel antisite defects can be reduced by, {\it e.g.}, synthesizing the material under the environments associated with points A and B in Fig.~\ref{fig;chempot}(b); however, even at these two points the formation energy of Ni$_{\rm Li}^{+}$ in \ce{LiNiO2} is still quite low (0.96 eV), unlike in \ce{LiCoO2} where the formation energy of Co$_{\rm Li}^{+}$ can be as high as 2.08 eV, {\it cf.}~Table~\ref{tab:formenergy}.

Let us now turn our discussion to the mechanisms for electronic and ionic conduction. The electronic or ionic conductivity $\sigma$ resulted from hopping of a defect X carrying charge $q$ can be defined as $\sigma = q m c$, where $m$ and $c$ are the defect's mobility and concentration, respectively.\cite{elliot} Let us assume that $c$ contains both thermally activated and athermal defect concentrations. The athermal defects can be, {\it e.g.}, pre-existing small polarons or lithium vacancies, or those polarons and vacancies formed during delithiation. Eqn.~(\ref{eq:concen}) can then be rewritten as
\begin{equation}\label{eq:concen2}
c=c_{a} + c_{0} \mathrm{exp}\left(\frac{-E^{f}}{k_{B}T}\right),
\end{equation}
where $c_{a}$ is the athermal defect concentration and $c_{0}$ is a prefactor. The mobility of the defects can also be assumed to be thermally activated, so
\begin{equation}\label{eq:mobility}
m=m_{0} \mathrm{exp}\left(\frac{-E_{m}}{k_{B}T}\right),
\end{equation}
where $m_{0}$ is a prefactor and $E_{m}$ is the migration barrier. When the athermal defect concentration is small, {\it e.g.}, at high temperatures and in nearly fully lithiated \ce{LiMO2}, the observed temperature dependence of the conductivity will be dominated by the second term in eqn.~(\ref{eq:concen2}) and shows an effective, intrinsic activation energy
\begin{equation}\label{eq;Ea} 
E_{a} = E^{f} + E_{m}, 
\end{equation} 
which includes both the formation energy and migration barrier. When the athermal defect concentration is large, {\it e.g.}, at low temperatures and/or in partially delithiated Li$_{x}$MO$_{2}$ ($x<1$), the contribution to the electrical conductivity will be dominated by the athermal term in eqn.~(\ref{eq:concen2}), and the activation energy will include only the migration part, {\it i.e.}, $E_{a}=E_{m}$. 

As mentioned in Sec.~3.3, our results indicate that intrinsic point defects cannot act as sources of band-like electrons and holes, and there are no (or negligible) free holes or electrons in \ce{LiMO2}. The electronic conduction thus has to proceed via hopping of $\eta^{+}$ and/or $\eta^{-}$. For the ionic conduction, lithium vacancies are most likely to be the charge-carrying defects, because other ionic defects either have very high formation energies and/or are expected to be immobile. From the calculated formation energies and migration barriers for the polarons and lithium vacancies listed in Table~\ref{tab:formenergy}, one can easily estimate the activation energies of the electronic and ionic conductivities using the above formulae for $E_{a}$. For example, the activation energy associated with $\eta^{+}$ in Li$_{x}$CoO$_{2}$ can be as low as 0.10 eV, which is the migration barrier of $\eta^{+}$; in nearly fully lithiated \ce{LiCoO2}, the intrinsic activation energy can be as low as 0.99 eV, which is the lowest calculated formation energy plus the migration barrier of $\eta^{+}$, {\it cf.}~Table~\ref{tab:formenergy}. We find that the contribution to the electronic conductivity from hopping of $\eta^{+}$ is almost always dominant, except at point D in Fig.~\ref{fig;chempot}(a) in the case of nearly fully lithiated \ce{LiCoO2} where the intrinsic activation energy associated with $\eta^{-}$ is lower than that associated  with $\eta^{+}$ by 0.40 eV. In \ce{LiNiO2}, $\eta^{+}$ and $\eta^{-}$ have comparable contributions to the electronic conductivity.

In nearly fully lithiated \ce{LiMO2}, the calculated formation energy of lithium divacancies is very high, therefore $V_{\rm Li}^{-}$ is expected to predominantly contribute to the ionic conductivity. The intrinsic activation energy associated with the diffusion of $V_{\rm Li}^{-}$ via a monovacancy mechanism can be as low as 1.25 eV (M=Co) or 1.43 eV (M=Ni), which is the lowest calculated formation energy value plus the migration barrier, {\it cf.}~Table~\ref{tab:formenergy}. In Li$_{x}$MO$_{2}$, on the other hand, the lithium vacancy concentration is high and vacancy agglomerates such as divacancies and trivacancies may become energetically favorable. The migration of lithium vacancies in this case is expected to occur via a divacancy mechanism, and the activation energy can then be as low as 0.18 eV (M=Co) or 0.26 eV (M=Ni), which is the calculated vacancy migration barrier, {\it cf.}~Fig.~\ref{fig;migration}.

\section{Conclusions}

We have carried out DFT studies of the bulk properties and intrinsic point defects in layered \ce{LiMO2}, using the HSE06 screened hybrid density functional. We find that stoichiometric \ce{LiCoO2} is stable in a large region in the chemical-potential diagram, whereas the stability region of Jahn-Teller distorted \ce{LiNiO2} is much smaller. \ce{LiNiO2} without the Jahn-Teller distortion is not stable toward competing Li$-$Ni$-$O phases.

\ce{LiCoO2} has a complex defect chemistry, resulting partly from the ability of Co ions to be stable in different charge and spin states. Different electronic and ionic defects such as small hole and electron polarons, lithium and transition-metal antisite defects, and lithium vacancies can form with high concentrations under different synthesis conditions. Cobalt antisites can be eliminated by synthesizing \ce{LiCoO2} under Li-excess (Co-deficient) environments. In the lithium over-stoichiometric \ce{LiCoO2}, negatively charged lithium antisites are charge-compensated by positively charged small (hole) polarons. Oxygen vacancies have high formation energies and are thus not likely to form in the interior of the material.

In Jahn-Teller distorted \ce{LiNiO2}, both small hole and electron polarons are always the dominant intrinsic point defects and have a low formation energy, indicating that a certain amount of Ni$^{3+}$ ions undergo charge disproportionation. Nickel antisites also have a low formation energy and hence high concentration. Our results suggest that tuning the synthesis conditions may lower the concentration of nickel antisites but would not remove the charge disproportionation.

Finally, we find that intrinsic point defects in layered oxides \ce{LiMO2} cannot act as sources of band-like electrons and holes, and the materials cannot be doped $n$- or $p$-type. The electronic conduction proceeds via hopping of small polarons, and the ionic conduction proceeds via migration of lithium vacancies through either a monovacancy or a divacancy mechanism. In \ce{LiCoO2}, the activation energy associated with hole polarons can be as low as 0.10 eV, and that associated with lithium vacancies can be as low as 0.18 eV. In \ce{LiNiO2}, the lower limit of the activation energy for hole polarons is higher (0.21$-$0.28 eV) partly because of associated Jahn-Teller fluctuations and 0.26 eV for migration of lithium vacancies.

\begin{acknowledgments}

We acknowledge helpful discussions with Noam Bernstein, and the use of computing facilities at the DoD HPC Center and the Center for Computationally Assisted Science and Technology (CCAST) at North Dakota State University. This work was supported, in part, by the U.S.~Department of Energy (Grant Nos.~DE-FG52-08NA28921 and DE-SC0001717), National Science Foundation EPSCoR (Award No.~EPS-0814442), and CCAST. Funding for M.D.J.~was provided by the U.S.~Office of Naval Research through the Naval Research Laboratory's Basic Research Program.

\end{acknowledgments}

%\bibliography{layeredoxides} %your .bib file
%

\end{document}